\documentclass[prl,aps,amssymb,twocolumn,superscriptaddress,showpacs]{revtex4}
\usepackage{psfig}
\usepackage{graphicx}
\def\ket#1{\vert#1\rangle}
\def\bra#1{\langle#1\vert}
\def\braket#1#2{\langle#1\vert#2\rangle}
\def\br{{\bf r}}
\def\bk{{\bf k}}

\def\re{{\rm Re}}
\def\pmb#1{\setbox0=\hbox{#1}%
\hbox{\kern-.025em\copy0\kern-\wd0
\kern.05em\copy0\kern-\wd0
\kern-0.025em\raise.0433em\box0} }
\def\taub{{\pmb{$\tau$}}}

\begin{document}
\title{Ballistic conductance of magnetic Co and Ni nanowires with ultrasoft
pseudo-potentials}
\author{Alexander Smogunov}
\affiliation{SISSA, Via Beirut 2/4, 34014 Trieste (Italy)}
\affiliation{INFM/Democritos National Simulation Center, Via Beirut 2/4, 34014
Trieste (Italy)}
\affiliation{Voronezh State University, University Sq. 1, 394006 Voronezh (Russia)}
\author{Andrea Dal Corso}
\affiliation{SISSA, Via Beirut 2/4, 34014 Trieste (Italy)}
\affiliation{INFM/Democritos National Simulation Center, Via Beirut 2/4, 34014
Trieste (Italy)}
\author{Erio Tosatti}
\affiliation{SISSA, Via Beirut 2/4, 34014 Trieste (Italy)}
\affiliation{ICTP, Strada Costiera 11, 34014 Trieste(Italy)}
\affiliation{INFM/Democritos National Simulation Center, Via Beirut 2/4, 34014
Trieste (Italy)}

\date{\today}

\begin{abstract} 
The scattering-based approach for calculating the ballistic conductance of
open quantum systems is generalized to deal with magnetic transition metals as
described by ultrasoft pseudo-potentials.
As an application we present quantum-mechanical conductance calculations
for monatomic Co and Ni nanowires with a magnetization reversal.
We find that in both Co and Ni nanowires,
at the Fermi energy, the conductance of $d$ electrons is blocked by a
magnetization reversal, while the $s$ states (one per spin) are perfectly
transmitted. $d$ electrons have a non-vanishing transmission in a small
energy window below the Fermi level. Here, transmission is larger in Ni
than in Co.
\end{abstract}
\maketitle

\section{I. Introduction}
Understanding electron conduction through atomic and molecular wires connecting two
macroscopic electrodes is presently a very active research topic. Conductance
measurements for metal nanocontacts display flat plateaus and abrupt drops during
the elongation attributed to atomic rearrangements.
In noble metals (Cu, Ag, Au) and alkali metals (Li, Na, K) the last conductance 
step before breaking, most likely corresponding to a monatomic nanocontact or
a nanowire, has a value $2G_0$ \cite{1} ($G_0=e^2/h$ is the conductance quantum per spin)
which can be rationalized with the free propagation of 2 valence $s$-electrons
(one per each spin). 
For magnetic transition metals such as Co and Ni the experimental data are less consistent.
Oshima {\it et al.}~\cite{oshima}, who worked in vacuum, at variable temperature,
and with the possibility of an external magnetic field, found conductance
steps in Ni nanocontacts preferentially near $2G_0$ and $4G_0$ at room temperature (RT) and zero field,
near $4G_0$ at 770 K and zero field, and near $3G_0$ (occasionally near $G_0$) at RT with a field.
Ono {\it et al.}~\cite{ono} reported again $2G_0$ for Ni in zero field, and $G_0$
for Ni in a field. Recently, Rodrigues {\it et al.}~\cite{ugarte2} observed 1 
conductance quantum at RT and zero field in a Co atomic chain.
Ruitenbeek and collaborators \cite{ruitenbeek} finally obtained low temperature zero field
data on a variety of magnetic metals, and reported a dominance of conductance steps
between $2G_0$ and $3G_0$ in Co and Ni.
Despite the poor agreement between these results it is clear that the last conductance step
is anomalously small with respect to the number of valence electrons determining the number
of conducting channels, since in Co and Ni there are many $d$ bands crossing the Fermi level.

Several theoretical methods are available to study the transport
properties of atomic scale conductors, each of them tailored to the
underlying electronic structure
scheme \cite{C7,C4,palacios,wavelet1,wavelet2,lang,hirose,finitediff,lkkr,kkr,wortmann}.
Among these are methods based on nonequilibrium Green's functions combined with a localized
basis set \cite{C7,C4,palacios} or with a system-independent wavelet basis set 
\cite{wavelet1,wavelet2}.
A scattering approach for conductance calculation has been applied by Lang and co-workers 
\cite{lang}, and Tsukada and co-workers \cite{hirose} using the jellium model for
the electrodes. 
A formulation based on the real-space finite-difference approach has been recently
proposed and applied to calculate conductance of gold and aluminum nanowires \cite{finitediff}.
Recently, the KKR method using the Baranger and Stone's
formulation of the ballistic transport has been presented \cite{kkr}.

In order to deal with magnetic transition metals
nanowires, we need a formulation capable of handling accurately the rather localized
$3d$ electrons. The approach to the problem of $d$ electrons
based on Vanderbilt ultrasoft pseudo-potentials (US-PPs)~\cite{vanderbilt}
has gained widespread use in recent years.
These PPs allow the accurate and efficient description
of large-scale systems, while
mantaining all the advantages of a plane waves basis set.
We already applied US-PPs to several electronic structure
and nanowire problems~\cite{dalcorso1}.
Recently, Choi and Ihm~\cite{choi} presented a method based on plane waves
to solve the tip-nanocontact-tip electron scattering problem 
in real atomic contacts and calculate from that
the ballistic conductance of open quantum systems via the Landauer-B\"uttiker formula. 
Their approach was originally formulated with norm-conserving 
(NC-PPs) Kleinman-Bylander-type pseudo-potentials; but that is not very suitable
for the localized $d$ electrons.
In the present paper a generalization of this method to US-PPs is presented.
As we shall show, it turns out that the solution
of the scattering problem is not more difficult
with US-PPs than with NC-PPs, while the calculation of the transmission 
coefficient requires a generalization of the expression for the probability 
current carried by a propagating state, which is put forward in this paper. 
The extra workload introduced by the US-PPs is in fact negligible.
Noting that a method of Thygesen {\it et al.} \cite{wavelet1}, formulated 
with a wavelet basis set, is also able to deal with US-PPs, we will
in Sec. IIIA  compare our results with those of Ref.\cite{wavelet1} 
in the specific example of an infinite monatomic Al wire with an H atom 
adsorbed on the side.

The paper is organized as follows. In Sec. II we describe a scattering-based 
approach for conductance calculation  with ultrasoft pseudo-potentials extending 
Choi and Ihm's method. In Sec. III we present some test calculations and apply 
our approach to study electron conduction in monatomic magnetic nanowires.  
A summary discussion is given in Sec. IV.

\section{II. Method}
We study electron transport in an open quantum system consisting of a scattering 
region ($0<z<L$) attached to the left ($z<0$) and to the right ($z>L$) to 
semi-infinite generic electrodes. We assume that the
electrons move ballistically in the self-consistent potential with reflection 
and transmission restricted to the scattering region. The self-consistent potential is 
obtained performing ground state density functional theory (DFT)
calculations with a supercell containing the scattering region and some portion 
of left and right electrodes. In the $xy$ plane the system is repeated 
periodically and in this plane the scattering states propagating at the 
energy $E$ have the Bloch form and can be classified with a $\bk_\perp$ index.
Different ${\bf k}_\perp$ do not mix and can be treated separately.
Furthermore we consider sufficiently large supercells in the $xy$ plane
and limit the calculation to the two-dimensional $\Gamma$ point, $\bk_\perp=0$.
The magnetic properties of the system are treated within
the local spin-density approximation (LSDA). Therefore electrons with different spin
polarization move independently in different self-consistent potentials and 
the scattering problem is solved separately for the two spin directions.
Therefore we will from now on solve the problem at some fixed ${\bf k}_\perp$ 
and spin index. Our approach and calculations will neglect all spin fluctuations,
of either quantum or thermal origin, assuming a statically spin polarized state
of the nanocontact.  

We describe transition metals atoms such as Co and Ni by ultrasoft 
pseudopotentials (US-PPs) \cite{vanderbilt}.
In that scheme the scattering state $\Psi$ at the energy $E$ 
is a solution of single-particle Kohn-Sham equation
\cite{laasonen} 
(atomic units $\hbar=2m=1,~e^2=2$ are used): 
\begin{equation} 
E\hat S \ket{\Psi}=\left[-\nabla^2+V_{\rm eff}+ 
\hat V_{NL}\right] \ket{\Psi}, 
\label{upeq1}
\end{equation} 
where $V_{\rm eff}$ is the effective local potential 
(see Refs.\cite{vanderbilt,laasonen}) and 
$\hat V_{NL}$ is the nonlocal part of US-PP: 
\begin{equation} 
\hat V_{NL}=\sum_{Imn}D^I_{mn}\ket{\beta^I_m} 
\bra{\beta^I_n}, 
\end{equation} 
constructed using the set of projector functions $\beta^I_m$ associated 
with atom $I$. The functions $\beta^I_m$ are localized within spheres of radius $r_c$
centered at the atom $I$. The coefficients $D^I_{mn}$ characterize
the US-PP and depend also on the effective potential $V_{\rm eff}$ \cite{laasonen}. 
The main difference with respect to NC-PPs is the appearance in the lhs of Eq.(\ref{upeq1}) 
of the overlap operator: 
\begin{equation}
\hat{S}=1+\sum_{mn}q_{mn}^I\ket{\beta_m^I}\bra{\beta_n^I}
\end{equation}
as a consequence of the relaxation of norm-conserving condition.
Here the coefficients $q^I_{mn}$ are the integrals of 
the augmentation functions defined in Refs.~\cite{vanderbilt,laasonen}.
Since the energy $E$ is a fixed input parameter one can rewrite the Eq.(\ref{upeq1})
as following:
\begin{equation} 
E\Psi(\br)=\left[-\nabla^2+V_{\rm eff}\right]\Psi(\br)+ 
\sum_{Imn}\tilde D^I_{mn}\braket{\beta^I_n} 
{\Psi}\beta^I_m(\br-{\bf R}_I)
\label{upeq2} 
\end{equation} 
with $\tilde D^I_{mn}=D^I_{mn}-Eq^I_{mn}$, thus recovering a form similar to the case of
NC-PPs.

As mentioned above, due to the translational symmetry the scattering states 
have the usual Bloch form in the $xy$ plane:
\begin{equation}
\Psi(\br_{\perp}+{\bf R}_\perp,z)=e^{i{\bf k}_\perp
{\bf R}_\perp} \Psi(\br_\perp,z),
\label{bloch}
\end{equation}
where ${\bf R}_\perp$ are the lateral supercell lattice vectors.

\subsection{A. Electrode region}
Deep within the two electrodes ($z<0,~z>L$) the scattering state $\Psi$ originating from
the rightward propagating Bloch wave $\psi_j$ of the left electrode
has an asymptotic form:
\begin{equation}
\Psi= \left\{
\begin{array}{ll}
\psi_j+\sum\limits_{i\in L}r_{ij}
\psi_{i}
, & \quad z<0
\\
\quad\sum\limits_{i\in R}t_{ij}
\psi_{i}
, & \quad z>L
\end{array}
\right.
\label{asymptotic}
\end{equation}
where summation over $i\in L~(i\in R)$ includes the generalized Bloch states
in the left (right) electrode at the energy $E$ which propagate or decay to the left
(right). The generalized Bloch states (both propagating and decaying)
constitute the so-called complex band structure of a solid~\cite{choi}.
They have the same Bloch form (\ref{bloch}) in the $xy$ plane while 
along the $z$ axis, where the translational symmetry is absent, 
they satisfy a condition:
\begin{equation}
\psi_j(\br_{\perp},z+d)=e^{i kd}\psi_j(\br_{\perp},z),
\label{perz}
\end{equation}
where $k$ is in general a {\it complex} number and $d$ is the length of the unit cell in the $z$ 
direction of the corresponding bulk crystal. The wave functions $\psi_j$ with 
${\rm Im}k=0$ and ${\rm Im}k\neq0$ describe propagating and decaying (or growing) states, 
respectively. In the bulk, only the former are of course eigenstates.

Inside the unit cell of the electrode $z_0<z<z_0+d$ the general solution 
of the integro-differential Eq.(\ref{upeq2}) can be written as a linear combination \cite{choi}:
\begin{equation} 
\psi_j(\br)=\sum_n c_{n,j}\phi_n(\br)+\sum_{Im} c_{Im,j} 
\phi_{Im}(\br). 
\label{expansion1} 
\end{equation} 
Here $\phi_n$ are linearly independent solutions of the homogeneous equation: 
\begin{equation} 
E\phi_n(\br)=\left[-\nabla^2+V_{\rm eff}(\br)\right]\phi_n(\br), 
\label{phin}
\end{equation} 
and $\phi_{Im}$ is a particular solution of the inhomogeneous equation: 
\begin{eqnarray} 
E\phi_{Im}(\br)&=&\left[-\nabla^2 
+V_{\rm eff}(\br)\right]\phi_{Im}(\br) \nonumber\\
&+&\sum_{{\bf R}_\perp} 
e^{i{\bf k}_\perp {\bf R}_\perp} 
\beta^I_m(\br-\taub^I-{\bf R}_\perp),
\label{phii}
\end{eqnarray} 
where $\taub^I$ is the vector-position of the $I$-th atom in the unit cell.
Both $\phi_n$ and $\phi_{Im}$ are $(x,y)$ periodic as in
Eq.~(\ref{bloch}) and the summation over $Im$ in Eq.~(\ref{expansion1}) involves 
all the projectors in the unit cell. We calculate the functions $\phi_n$ 
and $\phi_{Im}$ following the method described in Ref.\cite{choi}.

The function (\ref{expansion1}) is the solution of Eq.(\ref{upeq2}) 
if the coefficients $c_{Im,j}$ are given by: 
\begin{equation} 
c_{Im,j}=\sum_n \tilde D^I_{mn} 
\braket{\beta^I_n}{\psi_j}. 
\label{cim1}
\end{equation} 
The allowed values of $k$ for a given energy $E$ can be determined by 
imposing Eq.(\ref{perz}) along $z$ to the function 
$\psi_j$ and to its $z$-derivative: 
\begin{equation} 
\psi_j(\br_\perp,z_0+d)=e^{i 
k 
d} 
\psi_j(\br_\perp,z_0),
\label{perz1}
\end{equation} 
\begin{equation} 
\psi'_j(\br_\perp,z_0+d)=e^{i 
k 
d} 
\psi'_j(\br_\perp,z_0).
\label{perz2}
\end{equation} 
Inserting Eq.~(\ref{expansion1}) into Eqs.~(\ref{cim1})-(\ref{perz2}) one can show
that the last three equations are equivalent to the
generalized eigenvalue problem: 
\begin{equation}
AX=e^{i 
k 
d}BX. 
\end{equation} 
where $A$ and $B$ are matrices.
We solve this problem to obtain in general a complex $k$ and coefficients 
$X=\Big\{c_{n,j},c_{Im,j}\Big\}$ determining the generalized Bloch state 
$\psi_j$ at a given energy $E$ and $\bk_\perp$.

\subsection{B. Scattering region}
Inside the scattering region ($0<z<L$) the scattering state $\Psi$
can be represented in a form similar to Eq.(\ref{expansion1}):   
\begin{equation} 
\Psi(\br)=\sum_n c_n\phi_n(\br)+\sum_{Im} c_{Im} 
\phi_{Im}(\br), 
\label{expansion2}
\end{equation} 
where the functions $\phi_n$ and $\phi_{Im}$ are the solutions of Eqs.(\ref{phin},\ref{phii})
in the region $0<z<L$ and the summation over $Im$ involves all the projectors within
the scattering region. The scattering state $\Psi$ is completely determined 
by coefficients \{$c_n, c_{Im}, r_{ij}, t_{ij}$\}. The coefficients $c_{Im}$ are now determined by:
\begin{equation} 
c_{Im}=\sum_n \tilde D^I_{mn} 
\braket{\beta^I_n}{\Psi}
\label{cim2}
\end{equation} 
There is one more set of equations on $c_{Im}$
for the nonlocal spheres intersecting the boundaries of the scattering region
and thus shared with the electrodes. One has:
\begin{equation} 
c_{Im,j}+\sum_{i}r_{ij}c_{Im,i}= 
\sum_n \tilde D^I_{mn} 
\braket{\beta^I_n}{\Psi}=
c_{Im} 
\label{crosl}
\end{equation} 
for spheres intersecting the plane $z=0$ and 
\begin{equation} 
\sum_{i} t_{ij}c_{Im,i}= 
\sum_n \tilde D^I_{mn} 
\braket{\beta^I_n}{\Psi}=
c_{Im} 
\label{crosr}
\end{equation}
for spheres intersecting the plane $z=L$.
Eqs.(\ref{cim2})-(\ref{crosr}) together with the usual matching conditions on the wave function and 
its $z$-derivative on the boundary planes $z=0$ and $z=L$ give a set of linear algebraic equations.
Solving this set one obtains the unknown coefficients \{$c_n, c_{Im}, r_{ij}, t_{ij}$\}.
The accuracy of the matching procedure described above was tested on various systems having
no scattering region (e.g., infinite monatomic wires). In such systems different propagating
modes do not mix and must have unit transmission probability ($t_{ij}=\delta_{ij}$).
We found this condition to be satisfied with a very high accuracy 
($|t_{ij}-\delta_{ij}|\sim 10^{-6}$).

\subsection{C. Conductance calculation}
The ballistic conductance $G$ in the linear regime (infinitely small applied voltage)
is related to the total transmission $T$ at the Fermi energy by the
Landauer-B${\rm \ddot{u}}$ttiker formula $G=G_0T$, where $G_0=e^2/h$ is the conductance
quantum per spin. The total transmission is given by:
\begin{equation}
T=\sum_{ij}|T_{ij}|^2={\rm Tr}[{\bf T}^+{\bf T}],
\label{tran}
\end{equation}
where {\bf T} is the matrix of normalized transmission amplitudes
$T_{ij}=\sqrt{I_i/I_j}\cdot t_{ij}$ and $I_j$ is the probability current of the
Bloch state $\psi_j$ in the $z$ direction. Note that only rightward propagating states
in both left and right electrodes should be considered so the matrix {\bf T}
is of dimensions $M_R\times M_L$ where $M_L$ and $M_R$ are the number
of propagating modes in the left and right electrodes, respectively.
The sum over $i,j$ is over the states of both spin polarizations present at
the chosen energy. In our LSDA scheme the two spin channels are decoupled,
and therefore the $T_{ij}$ is nonvanishing only if
$i$ and $j$ have the same spin. No spin flip is allowed in our calculation.

The eigenvectors of the Hermitian matrix ${\bf T}^+{\bf T}$ determine the coefficients
of a unitary transformation from the set of Bloch states $\psi_j$
to the conductance eigenchannels~\cite{channels}.
In the eigenchannel basis the matrix ${\bf T}^+{\bf T}$ is diagonal.
Calling $T_i$ the eigenvalues, the conductance is a sum of independent
contributions from each eigenchannel:
\begin{equation}
G=G_0\sum_i T_i,
\end{equation}
where $T_i$ gives the transmission probability for $i$-th eigenchannel.

In order to calculate the total transmission $T$ one needs to know the
current $I_j$ carried by the propagating Bloch state $\psi_j$ in the $z$ direction.
The $\psi_j$ are actually pseudo-wave functions, and coincide with
the all-electron wave functions only outside the core regions.
Therefore the usual expression for the current flowing through a plane $S$
perpendicular to the $z$ axis and located at $z_0$:
\begin{equation}
I^0_{j}=2{\rm Im}\left[\int_S\psi^*_j(\br_\perp,z_0){\partial \psi_j(\br_\perp,z)\over
\partial z}\bigg|_{z=z_0}
d^2\br_\perp\right]
\end{equation}
is valid only when the plane $z_0$ does not intersect the nonlocal spheres.
At a general point it must be modified as:
\begin{eqnarray}
I_j&=&I^0_j-2{\rm Im}\bigg[ \sum_{Imn}\tilde{D}^I_{mn}\braket{\beta^I_n}{\psi_j}\times \nonumber\\
&&\int_{-\infty}^{z_0}dz\int_S\beta^I_m(\br-{\bf R}_I)
\psi^*_j(\br)d^2\br_\perp\bigg].
\label{current}
\end{eqnarray}
This expression has been derived by requiring the two following properties.
First, the term added to $I^0_{j}$ guarantees that $I_j$ does not depend
on $z_0$ once the wave function $\psi_j$ satisfies Eq.(\ref{upeq2}) and
has the Bloch form in the $xy$ plane. Second, at all $z_0$ where the plane $S$
does not intersect the nonlocal spheres this additional term vanishes
so that $I_j=I^0_j$.
The accuracy of the current calculation with Eq.(\ref{current}) was tested by checking
the unitarity of the scattering matrix which is a consequence of current conservation.
For each propagating state $\psi_j$ one must have $\sum_i T_{ij}+\sum_i R_{ij}=1$, 
where $T_{ij}$ and $R_{ij}$ are the normalized transmission
and reflection coefficients, respectively. Again, this condition was 
found to be satisfied with very high accuracy for all the systems considered so far.

We point out that Eq.(\ref{tran}) assumes that the current operator is
diagonal in the basis of Bloch states. That is not always true.
A mixing might occur between degenerate states with the same $k$ in
$z$ direction (for example, states derived from atomic $p_x,p_y$
or $d_{xz},d_{yz}$ levels if the system has an axial symmetry around $z$ axis).
Therefore we must orthogonalize Bloch states with respect to the
current operator before applying Eq.(\ref{tran}) for the total transmission calculation.
The general off-diagonal matrix elements $I_{jk}$ of the current operator are given by:
\begin{eqnarray}
I_{jk}&=&I^0_{jk}+i\sum_{Imn}\tilde{D}^I_{mn}\times \nonumber\\
&&\hspace{-5mm}
\bigg[\braket{\beta^I_n}{\psi_k}\int_{-\infty}^{z_0}dz\int_S\beta^I_m(\br-\taub^I)
\psi^*_j(\br)d^2\br_\perp- \label{ijk}\\
&&\hspace{-3mm}
\braket{\beta^I_n}{\psi_j}^*\int_{-\infty}^{z_0}dz\int_S[\beta^I_m(\br-\taub^I)]^*
\psi_k(\br)d^2\br_\perp \nonumber
\bigg],
\end{eqnarray}
where
\begin{eqnarray}
I^0_{jk}&=&-i\int_S\bigg[\psi^*_j(\br_\perp,z_0){\partial\psi_k(\br_\perp,z)\over
\partial z}\bigg|_{z=z_0}- \\
&&\quad\quad\quad {\partial\psi^*_j(\br_\perp,z)\over
\partial z}\bigg|_{z=z_0}\psi_k(\br_\perp,z_0)\bigg]d^2\br_\perp.\nonumber
\end{eqnarray}

The ballistic conductance is calculated in three steps. First, we perform
the supercell DFT electronic structure calculation with the plane-wave code
({\tt PWscf})~\cite{pw} to obtain the self-consistent potential $V_{\rm eff}$ and the
screened coefficients $D^I_{mn}$. 
For spin-polarized systems both $V_{\rm eff}$ and $D^I_{mn}$ will depend on the
spin polarization. Second, we calculate the complex band structures of 
the infinite left and right electrodes and orthogonalize the propagating 
Bloch states with respect to the current operator using Eq.(\ref{ijk}).
The unit cells of the electrodes are chosen within the supercell,
beyond the scattering region, where the potential is bulk-like.
Finally, we calculate the transmission coefficients $t_{ij}$
for each rightward propagating state $\psi_j$ of the left electrode and obtain the total
transmission $T$ and eigenchannel transmissions $T_i$ as described above.
This procedure now represents a rigorous extension of Choi and Ihm's method to US-PPs.

\section{III. Test cases and application to monatomic magnetic nanowires}
We apply now the above scheme to calculate the ballistic conductance
for two test systems: a) a monatomic Al wire with an H impurity; b) carbon chains between
Al electrodes with a finite cross-section. The conductance of these systems has been
calculated by other methods and we can compare our results with the known data.
As our main application, we study the effect of the 
magnetization reversal on electron transport in monatomic Co and Ni nanowires (a case of
ballistic magnetoresistance). 

\begin{figure}
\hspace*{-0.8cm}
\includegraphics[width=9.7cm]{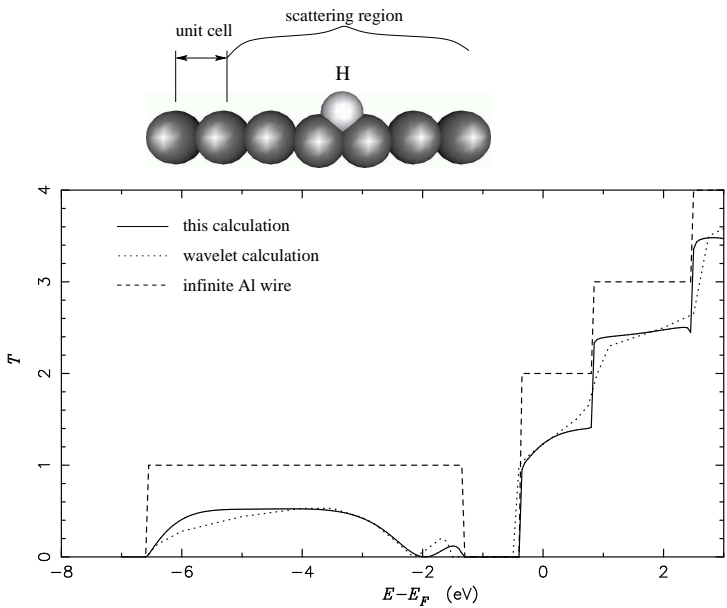}
\caption{\label{fig1} 
Transmission coefficient for monatomic Al wire with an H atom adsorbed on the side.
The corresponding result of Ref. [9] and the transmission coefficient of the perfect infinite
wire are shown by a dotted and a dashed line, respectively.
}
\end{figure}

\subsection{A. Monatomic Al wire with an H impurity}
We present first the conductance of a monatomic Al wire with an H atom adsorbed on the side 
and compare our result with the wavelet basis set calculations of Ref.\cite{wavelet1}. 
The supercell for electronic structure calculation along with its division onto the bulk-like
part (used for complex band structure calculations) and the scattering region 
are shown in Fig.~1. The LDA calculations were performed using the Perdew-Zunger 
parametrization \cite{perdew} of exchange-correlation energy.
The hydrogen atom was described by US-PP with the parameters given in Ref.\cite{dalcorso1}. 
Energy cutoffs of 25 and 250 Ry were used for the wave functions and for the charge density,
respectively. The BZ was sampled by three special k-points.   
The spacing between Al atoms was 2.39 \AA~while the wires were separated by $|R_\perp|=7.5$ \AA.
As in Ref.\cite{wavelet1} the positions of the adsorbate and of the two nearest Al atoms
were optimized while the other Al atoms in the supercell were kept at the positions 
corresponding to the perfect infinite wire. The relaxed Al-H distance was found to be 
1.82 \AA~ with the Al-H-Al angle of 79.1$^{\rm o}$.

In Fig.~1 one can see that the calculated transmission curve (solid line) is 
very similar to that of Ref.\cite{wavelet1} (dotted line). We recover the main 
features found in Ref.\cite{wavelet1}: a) the adsorption of H atom reduces the 
transmission coefficient by a factor approximately $0.5$ with respect to the 
transmission coefficient of the perfect infinite wire (which is merely determined 
by the number of propagating states at a given energy and
is shown by a dashed line in the figure); b) the drop of the transmission 
to zero at an energy about $-2.0$ eV. 
We note a small difference in the positions of the edges of transmission curves 
in two calculations (see, for instance, the low maximum at about $-1.5$ eV) 
which is a consequence of slightly different band structures of the 
bulk-like Al wires. A shift of the energy band leads to a corresponding shift 
of the transmission curve. Slight differences of transmission will also be related 
to the optimized atomic positions, presumably slightly different in our calculations 
and in those of Ref.\cite{wavelet1}. 

\subsection{B. Carbon chains}
As a second example we consider carbon chains attached to Al electrodes of finite cross-sections.
The electron conduction in such systems has been recently calculated using methods based on a
localized basis set \cite{C7,C4}. The system consists of a finite length carbon chain placed between
two Al electrodes of finite cross section oriented along Al(100) direction (see Fig. 2).
The carbon atoms are described by US-PPs with the parameters of Ref.\cite{dalcorso1}.
As in the previous case, the energy cutoffs of 25 and 250 Ry were used for 
the wave functions and for the charge density, respectively, and the BZ was 
sampled by three special k-points. The size of the supercell in the $xy$ plane was chosen to be
$L_x=L_y=12.15$ \AA. The electrode unit cell is composed of two layers and contains nine atoms.
The carbon atoms in the chain are separated by 2.5 a.u. The carbon chain ends are 
positioned in the hollow sites of the Al(110) electrode sufaces, the electrode-chain 
distance denoted as $D$.

\begin{figure}[b!!]
\hspace*{-0.5cm}
\includegraphics[width=9.4cm]{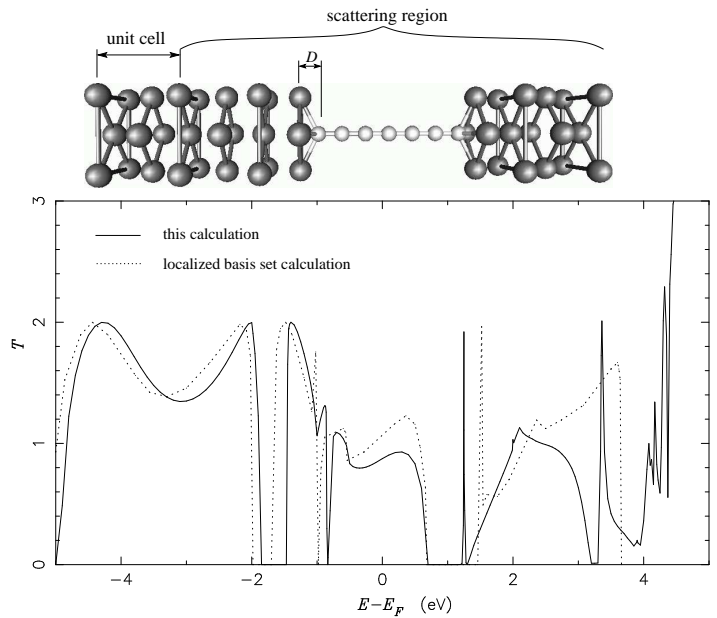}
\caption{\label{fig1}
Transmission coefficient for the seven-atom carbon chain between two finite cross-section Al electrodes
with the electrode-chain separation $D=1.0$ \AA.
The dashed line shows the corresponding result obtained with a localized basis set [6].
}
\end{figure}

We considered seven-atom (with $D=1.0$ \AA) and four-atom (with $D=1.9$ \AA) 
carbon chains previously studied with a localized basis set in Ref.\cite{C7} and 
Ref.\cite{C4}, respectively. Transmission coefficients calculated for these systems 
are presented in Figs. 2 and 3. Our transmission curves (solid lines) show qualitatively 
the same behaviour as the corresponding curves of Refs.\cite{C7},\cite{C4} 
(dashed lines). 
Broad maxima in the transmission coefficient are related
to the energy position of the carbon $\pi$ states while the energy windows with
vanishing transmission (e.g., energy intervals [$-1.75,-1.5$] and [0.7,1.2]) correspond 
to gap energies where no coupling exists between carbon chain and electrode 
states. We note some small discrepancies between the present plane waves calculation
and the localized basis set calculations, mainly concerning the positions of the 
edges of transmission curves. For example, our calculations give zero transmission 
coefficient in the energy window [$-1.75,-1.5$] to be compared
with the energy interval [$-1.9,-1.7$] of Ref.\cite{C7}. As before, the 
shifts of transmission curves appear to be related to slight differences in the 
band structures of the electrodes obtained using two different basis sets. 

\begin{figure}
\hspace*{-0.3cm}
\includegraphics[width=9cm]{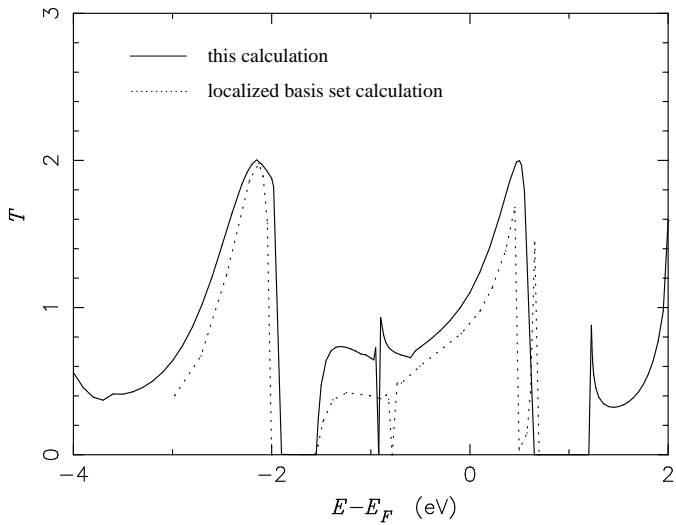}
\caption{\label{fig1}
Transmission coefficient for the four-atom carbon chain between two finite cross-section Al electrodes
as in Fig.~2 but with an electrode-chain separation $D=1.9$ \AA.
The dashed line shows the corresponding result obtained with a localized basis set [7].
}
\end{figure}

\subsection{C. Monatomic magnetic nanowires}
We now apply the method to calculate the ballistic conductance of 
monatomic Co and Ni nanowires with a single spin reversal.
We discuss first the electronic and geometrical structure of an infinite 
monatomic Co wire without spin reversal~\cite{parameters}.
Similar results for the Ni nanowire were reported in Ref.~\onlinecite{our1}.
The total energy and the magnetic moment of the Co wire as a function 
of the interatomic spacing are shown in the insets of Fig.4.
The ground state of the wire is ferromagnetic with an energy gain of about 
0.41 eV/atom with respect to the nonmagnetic wire. The equilibrium 
interatomic distance for the Co wire (about $a=2.11$~\AA) is similar to 
that of the Ni wire~\cite{our1} while the magnetic moment is more than twice 
($\mu=2.16~\mu_{\rm B}$/atom compared to $1.11~\mu_{\rm B}$/atom for the Ni wire).
Similar to the Ni wire~\cite{our1} the magnetic moment of the Co wire 
vanishes and the wire turns nonmagnetic in the unphysical region where 
the interatomic distance is taken less than approximately $1.6$~\AA\ (1.9~\AA~for Ni).

For a perfect infinite monatomic wire the number of conducting 
channels is just the number $N$ of bands crossing the Fermi level.
Since in the absence of tip-wire junction each channel has a unit 
transmission probability the ballistic conductance as given by the
Landauer-B${\rm \ddot{u}}$ttiker formula is $G=G_0N$. In Fig.4 we show 
the complex band structure of a monatomic Co wire obtained at the 
equilibrium interatomic spacing - an ingredient that will be needed 
for further conductance calculation. Complex bands for the majority 
(spin up) and minority (spin down) spin polarizations are displayed 
in Figs.~4a and 4b respectively. Qualitatively, the complex bands of the Co wire
look quite similar to those for a Ni wire~\cite{our2} though the exchange splitting
is twice as large (approximately 2 eV versus 1 eV for the Ni wire) which is
directly related to a stronger Hund's rule coupling and a higher magnetic 
moment per atom of the Co wire. As for the number of conductance channels,
from the figure one can count one spin up and six spin down (real) bands 
crossing the Fermi level, giving altogether seven propagating channels. 
The same result was previously found for the Ni wire. \cite{our1,our2}
Not surprisingly, this overall number of channels is much larger than 
any of the observed experimental conductance steps, of both nickel
and cobalt nanocontacts \cite{oshima,ono,ugarte2, ruitenbeek}.

\begin{figure*}
\includegraphics[width=14cm]{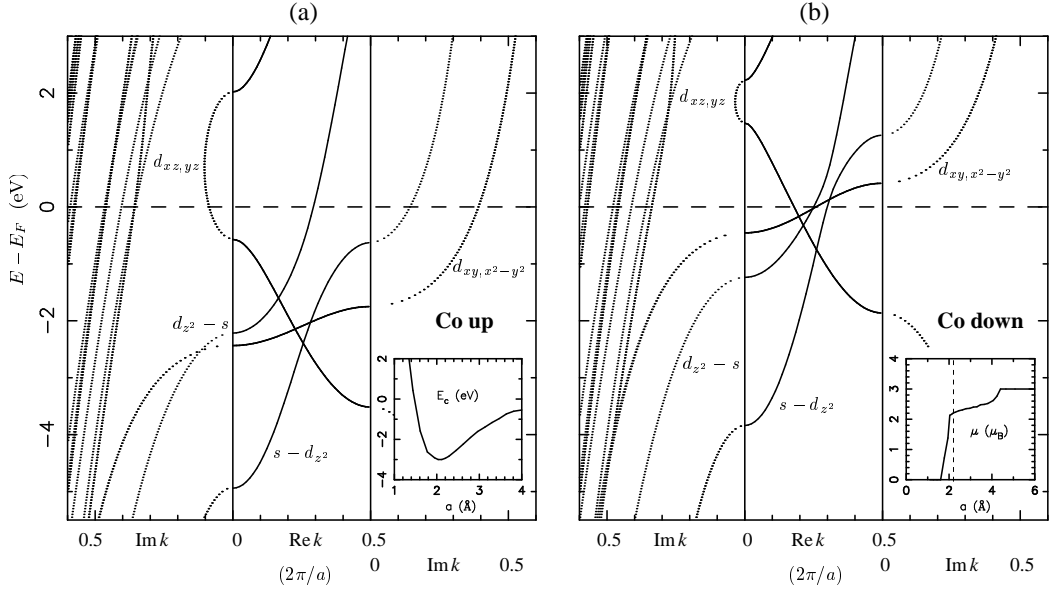}
\caption{\label{fig1}
Complex band structure of monatomic Co nanowire
for the majority (spin up) and minority (spin down) states.
Real bands, complex bands with $\re\ k=0$, and $\pi/a$
are plotted in the middle, left, and right panels, respectively.
Each band is labeled by its main atomic character. The total energy and
the magnetic moment per atom as a function of interatomic spacing are shown
on the left and right insets, respectively.
}
\end{figure*}

We consider now a monatomic nanowire with positive magnetization as our
left electrode, and another with negative magnetization as our right electrode,
the nanocontact thus consisting of a magnetization reversal.
We obtain the optimal shape of magnetization reversal by LSDA calculations,
minimizing the total energy of a supercell containing 12 atoms.
The detailed shape of the optimal magnetization reversal was obtained by
allowing magnetization to vary freely in magnitude and sign, while constraining 
the overall total magnetic moment to be zero (for computational details see Ref.~\onlinecite{our1}).
The resulting spin reversal turned out to be quite sharp and the 
ground state had, roughly speaking, 6 atoms with positive and 6 atoms with 
negative magnetization (see Fig.~5) While this way to build a collinear magnetization
reversal would be of course incorrect for a bulk Bloch wall, it may not be
an unreasonable description of magnetization reversal inside a very short nanocontact.
Bruno \cite{bruno} showed that the width of a nanoconstrained magnetic domain wall 
is essentially determined by the size of the constriction and can thus be 
much smaller than the (noncollinear) Bloch wall in 3D crystals or N\'eel walls in a film.
Essentially, in such an atomically constrained system, the domain wall
spin rotation must take place in such small distance to justify a spin collinear 
approximation.

The self-consistent potential for the nanowire with a spin reversal is subsequently used to solve 
the scattering problem. The scattering region and the unit cell of the left 
and right side of the wire are shown in Fig.~6. For symmetry reasons the total 
transmission is the same for electrons of both spin directions (this conclusion 
has also been checked numerically). Therefore we choose to consider only 
electrons with spin down polarization incident from the left on the
spin reversal region.

\begin{figure}[b!!]
\includegraphics[width=9cm]{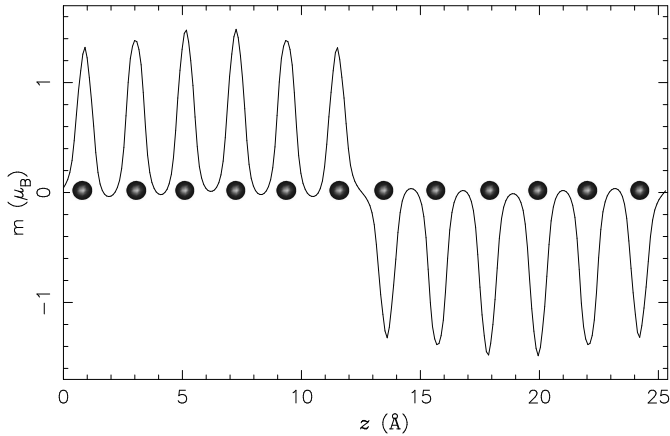}
\caption{\label{fig1}
Planar average of the magnetization along the Co nanowire.
}
\end{figure}

\begin{figure*}
\includegraphics[width=11cm,angle=-90]{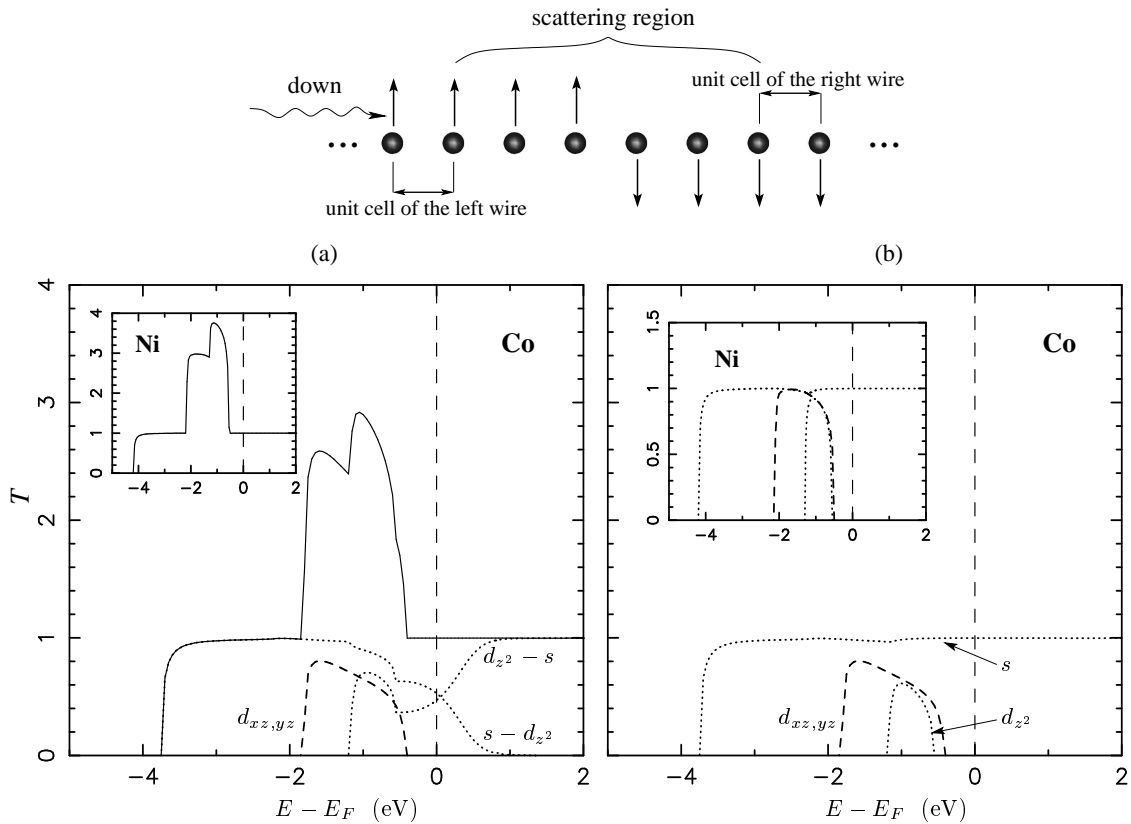}
\caption{\label{fig1}
Transmission coefficients for spin down electron incident from the left on a magnetization reversal in a monatomic
Co nanowire:
a) total transmission (solid curve) and contributions from each band;
b) eigenchannel transmissions $T_i$.
The corresponding data for a monatomic Ni nanowire are shown on the insets.
The bands and conductance channels are labeled by their main atomic character.
The Fermi energy is chosen to be zero (dashed vertical lines).
}
\end{figure*}

In Fig.~6a we present the total transmission (for one spin channel) and the 
contributions from each band as a function of energy for the Co wire with a 
spin reversal. The propagating state in the left half of wire can propagate
through the spin reversal and contribute to the total transmission only if it 
can be matched with an appropriate propagating state to the right. For 
example, at the Fermi energy there are 6 propagating states in the left side. 
However, only $s-d_{z^2}$ and $d_{z^2}-s$  states have nonzero transmission since
there is only one $d_{z^2}-s$ propagating state in the right side. The $d$ states
in particular can be matched only with decaying states in the right side and 
are therefore completely reflected due to the spin reversal. At energy about 
$E=E_F-0.4$ eV the two $d_{xz,yz}$ states start contributing to the transmission 
because there appear two states of the same symmetry in the right side of the wire.
On the contrary, the $d_{xy,x^2-y^2}$ states cannot get through the spin reversal 
at any energy and never contribute to conductance. Fig.~6b shows the eigenchannel 
transmissions $T_i$. The $s-d_{z^2}$ and $d_{z^2}-s$ states
are combined to form a single perfectly propagating conductance eigenchannel 
(essentially $s$ like). Another channel (mostly $d_{z^2}$), orthogonal to the 
first one, appears at energy about 0.6 eV below the Fermi energy. It is quite 
narrow and has a smaller transmission probability (about 0.6 at the maximum).
The two $d_{xz,yz}$ states form two eigenchannels of the same symmetry. They contribute
to the total transmission between 0.4~eV and 1.8~eV below the Fermi energy 
and are partially reflected (the transmission probability is about 0.8 at the maximum).

In the left and right insets of Fig.~6 we show the total transmission (for 
one spin channel) and eigenchannel transmissions, respectively, similarly calculated 
for a Ni nanowire with a spin reversal. One can see that again there is a single 
conductance channel perfectly propagating at all the energies studied
contributing one conductance quantum (per spin) to the total conductance at the Fermi energy.
However, the contribution from other $d$ channels at lower energies is now much larger
(at the maxima these channels have almost unit transmission probabilities).
This seems reasonable since the Co atoms in the wire have a larger
exchange splitting, and are therefore more reflecting, compared to the Ni atoms 
(as discussed above). 

Adding the transmission of the two spin channels at the Fermi energy we get 
the overall value of $G_0+G_0$ 
for the conductance. This is to be compared with $G_0+6G_0$ of the ferromagnetic wire.
Such a large difference between conductances of uniform ferromagnetic wire 
and the same ferromagnetic wire with a spin reversal is an example 
of magnetoresistance, analogous to that discussed earlier.\cite{garcia} 
Note that the minimal conductance value $2G_0$ obtained in presence of
spin reversal is still larger than that of ``fractional peak'' at approximately $G_0$ 
reported for Co and Ni nanowires~\cite{ugarte2}.
We should stress, however, that
the exact equivalence of the $s$ channels for the two spin polarizations,
leading to the factor two in our result, is only valid for our chosen
spin-symmetric geometry. The investigation of more complex magnetic and structural geometries
which might have lower conductance values will be the subject of future work.
The role of fluctuations should also be considered, as they
are likely to be important in a nanomagnetic system, particularly at high
temperature.

\section{IV. Conclusions}
We have generalized to magnetic transition metals the approach of Choi 
and Ihm for calculating the ballistic conductance of an open quantum system. 
The method has been implemented with ultrasoft pseudopotentials and plane wave 
basis set in a DFT-LSDA ab-initio scheme. After checking the results against
previous calculations based on different methods, we applied our method 
to calculate the ballistic conductance of monatomic Co and Ni nanowires with 
a single spin reversal. The $s$-like channel has a unit transmission probability 
at all energies for both Co and Ni nanowires. On the contrary, the transmission 
probability of the heavy $d$ electrons is heavily influenced by the spin reversal. 
For both wires we found that at the Fermi energy there is only one conductance 
channel per spin ($s$ like) while all $d$ electron 
ballistic conductance is completely blocked by the spin reversal, as was
previously guessed from band structure calculations \cite{our1}. 
In the energy window between 0.4~eV and 1.8~eV below the Fermi level the $d$ electron
conductance is smaller for the Co wire, due to an exchange splitting almost twice 
as large as that of the Ni wire. No structural distortions of linear 
chains (like zigzag structures or dimerizations) were considered in this paper,
as they are not likely to be relevant in the short nanocontact connecting
the two tips. More efforts to reproduce the realistic geometries of the
nanocontacts will form the subject of future work.

\section{Acknowledgments}
This work was partly sponsored by MIUR COFIN03 and FIRB RBAU01LXSH and
FIRB RBAU017S8R, by INFM (section F, G,
``Iniziativa Trasversale calcolo parallelo''), and by EU Contract ERBFMRXCT970155 (FULPROP).
Calculations were performed with the {\tt PWscf} package~\cite{pw}
on the IBM-SP4 at CINECA, Casalecchio (Bologna).


\begin{thebibliography}{28}
\bibitem [23] {parameters}
The LSDA calculations were performed with the Perdew-Zunger parametrization \cite{perdew} of exchange-correlation
energy. The core radii (in a.u.) of Co pseudo-potential are: $r_s^c=(2.0,2.2)$, $r_p^c=(2.2,2.3)$, $r_d^c=(1.6,2.2)$.
The kinetic energy cut-offs of the plane wave basis set were 25 Ry
and 300 Ry for the wave functions and the charge density, respectively.
$\bk$-point sampling is done with a uniform mesh
of 40 $\bk$-points in the 1D BZ zone.
The integration up to the Fermi level was done by
a standard broadening technique \cite{smearing} with the smearing
parameter of 0.005 Ry.
\end{thebibliography}
\end{document}